\documentclass[twocolumn,showpacs,aps,prb,preprintnumbers,amsmath,amssymb,floatfix]{revtex4}

\usepackage{graphicx}
\usepackage{dcolumn}
\usepackage{bm}
\usepackage{epsfig}

\begin{document}
\date{\today}

\title{ 
Helical Andreev bound states and superconducting Klein tunneling \\ in topological insulator Josephson junctions
}

\author{ G. Tkachov and E. M. Hankiewicz }

\affiliation{ 
Institute for Theoretical Physics and Astrophysics, University of W\"urzburg, Am Hubland, 97074 W\"urzburg, Germany,\\ 
}

\begin{abstract}
Currently, much effort is being put into detecting unconventional $p$-wave superconductivity in Josephson junctions based on topological insulators (TIs).
For that purpose we propose to use superconducting Klein tunneling, i.e., the reflectionless passage of Cooper pairs through a potential barrier 
in a gated ballistic junction. This phenomenon occurs due to the fact that the supercurrent is carried by helical Andreev bound states (ABSs) 
characterized by spin-momentum locking similar to the normal-state carriers.  
We derive the spectrum of the helical ABSs and the corresponding Josephson current for a junction made on the surface of a three-dimensional TI. 
The superconducting Klein tunneling is predicted to yield a non-sinusoidal current-phase relation and an anomalous critical current $I_c$ 
that does not vanish with increasing barrier strength. We also analyze the dependence of the $I_cR_n$ product 
(where $R_n$ is the normal-state junction resistance) on the microscopic parameters of the superconductor/TI interface, 
which leads to lower $I_cR_n$ values than expected from previous models of the proximity-effect Josephson junctions.  
\end{abstract}
\maketitle

\section{Introduction}
\label{Intro}

Unconventional $p$-wave superconductivity and Majorana states predicted in topological insulator (TI)/superconductor (S) junctions \cite{Fu08} 
have become a topic of vigorous research (see e.g. reviews \cite{Hasan10,Qi11,Tanaka12,Alicea12,Beenakker13a,GT13}).
In three-dimensional (3D) TIs the superconductivity can be induced 
in the proximity to a conventional superconductor (e.g., Al, W, or Nb) deposited
on the surface of the TI material.~\cite{D_Zhang11,Koren11,Sacepe11,Veldhorst12,Wang12_STI,Williams12,Maier12,Cho12,Oostinga13}
The unconventional superconductivity originates from the broken spin rotation symmetry of the helical surface states, 
allowing for mixed singlet $s$-wave and triplet $p$-wave pair correlations,
whereby the $p$-wave component inherits the spin-momentum locking of the surface states. 
Models for the superconducting proximity effect in the TIs have been proposed in Refs. 
\onlinecite{Stanescu10,Potter11,Labadidi11,Khaymovich11,Virtanen12,Yokoyama12,Black12},\onlinecite{Maier12},\onlinecite{GT13}, 
including the case of disordered TIs in which the $p$-wave superconductivity is suppressed.~\cite{GT13b}

Due to the induced $p$-wave superconductivity, the TI Josephson junctions are expected to support gapless Andreev bound states (ABSs)~\cite{Fu08} 
which have been studied theoretically in a variety of situations 
(see Refs.~\onlinecite{Fu09,Tanaka09,Linder10,Badiane11},\onlinecite{Olund12,Yamakage12},\onlinecite{GT13,Beenakker13b,Zhang13,Wieder13,Snelder13,Potter13}).    
Transport mediated by the gapless ABSs may therefore serve as evidence for the unconventional superconductivity in the TIs. 
Among such transport phenomena proposed in literature on TIs 
are the $4\pi$-periodic Josephson effect~\cite{Fu09,Yamakage12,Beenakker13b,Zhang13} and its signatures in the current noise,~\cite{Badiane11}
the magnetically controlled Josephson effect,~\cite{Tanaka09,Zhang13,Snelder13}
the phase-dependent electric conductance and noise~\cite{Wieder13}
as well as the anomalies in the current-phase relation and Fraunhofer pattern.~\cite{Potter13}

\begin{figure}[b]%
\begin{center}  
\includegraphics[width=1\linewidth]{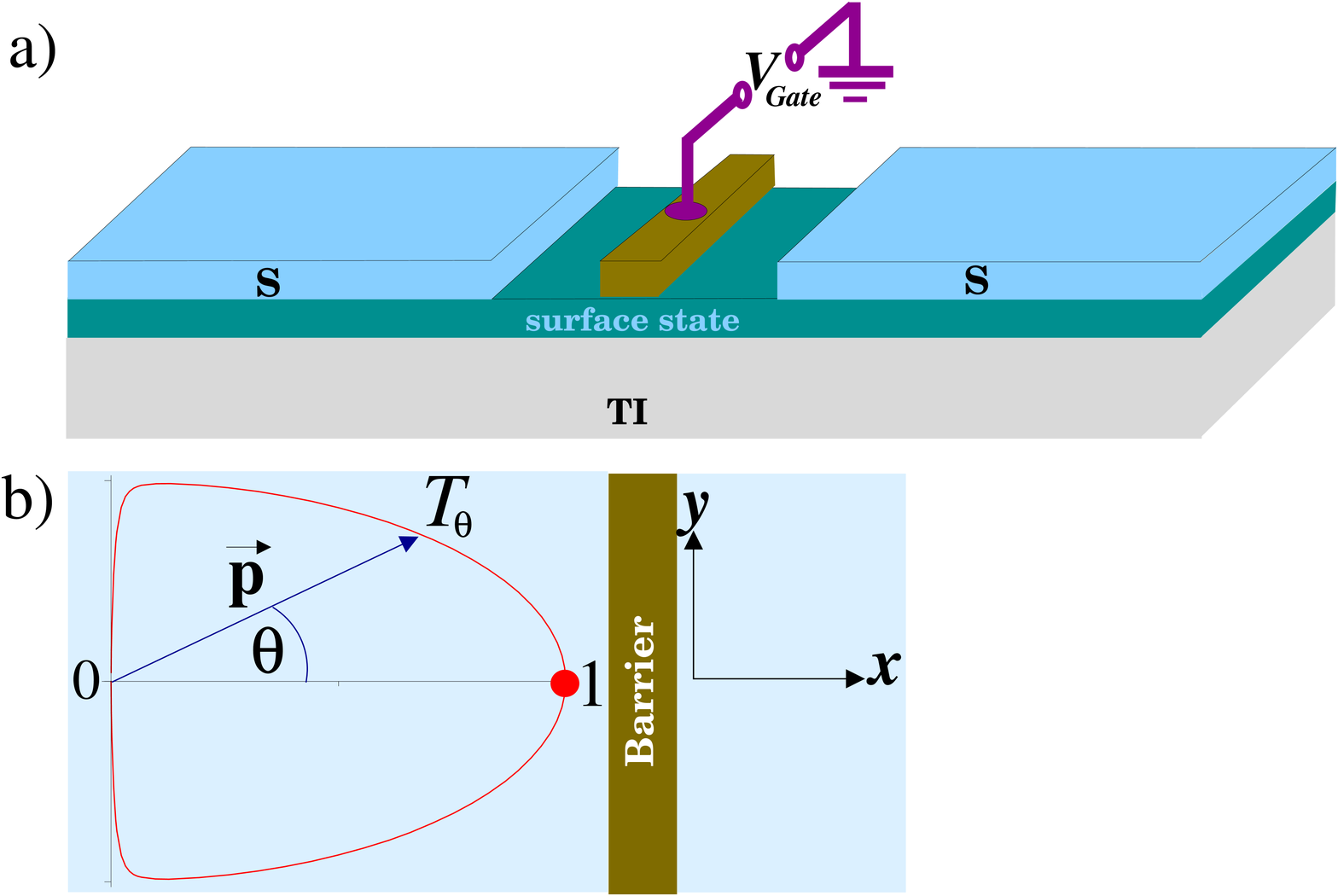}%
\end{center}
\caption{%
(a) Schematic of S/TI/S junction with a gated-induced barrier in the middle (see also the text). 
(b) Scattering geometry in junction plane and polar plot of barrier transparency $T_\theta$ as a function of incidence angle $\theta$ [see Eq. (\ref{T})]. 
}
\label{STIS}
\end{figure}

Based on our recent proposal,~\cite{GT13} in this paper we study another transport manifestation of the induced $p$-wave superconductivity in TIs, 
a phenomenon analogous to Klein tunneling in normal helical metals. 
It should be observable in ballistic TI Josephson junctions with a gated weak link [see Fig. \ref{STIS}(a)], 
which we model by introducing a potential barrier between the S terminals.    
We show that the induced $p$-wave superconductivity leads to helical ABSs that exhibit spin-momentum locking and 
depend strongly on the momentum direction with respect to the barrier. 
For the normal incidence at the barrier with $\theta =0$ [see Fig. \ref{STIS}(b)], 
the ABSs are protected by time-reversal symmetry and are gapless. 
In contrast, for the oblique incidence with $\theta \not =0$ there is no protection since 
the incident and reflected trajectories are not connected by time-reversal symmetry. 
In this case, the ABSs acquire a minigap $\propto \sqrt{ 1 - T_\theta }$ 
depending on the normal-state barrier transparency $T_\theta$ [see Fig. \ref{STIS}b].
As a consequence, for the normal propagation the supercurrent carried by the gapless ABSs is just the same as in the absence of the barrier. 
In this sense, there is a direct analogy with the normal-state Klein tunneling. 
Although the net supercurrent contains the contributions of all possible propagation directions, 
we demonstrate that the superconducting Klein tunneling is still identifiable by two features: 
the non-sinusoidal current-phase relation and the anomalous critical current $I_c$ that does not vanish with increasing barrier strength.   
We also analyze the $I_cR_n$ product (where $R_n$ is the normal-state junction resistance) as a function of the barrier strength and  
the microscopic parameters characterizing the S/TI interface.

The paper is organized as follows. In Sec. \ref{STI} we review the superconducting proximity effect in a 3D TI, 
emphasizing its two characteristic features: helical Bogolubov quasiparticles and the mixed $s$- and $p$-wave pair correlations. 
In Sec. \ref{ABS} we derive the helical ABSs and discuss their anisotropic properties. 
Section \ref{Current} presents the results for the current-phase relation, critical current, and the $I_cR_n$ product.
Section \ref{Summary} summarizes our results.

\section{ Induced superconductivity in TI$s$}
\label{STI}

\subsection{Model}
\label{Model}

We begin by setting up a microscopic model for the proximity effect in an infinite planar contact between a singlet $s$-wave superconductor (S) 
and a 3DTI. Assuming tunneling coupling between the systems, we will follow the general idea of McMillan's approach, 
\cite{McMillan68} further adapted to various low-dimensional systems in 
Refs. \onlinecite{Golubov04,GT04,GT05,Fagas05},\onlinecite{Stanescu10,Potter11},\onlinecite{Kopnin11,Stanescu11}. 
In this approach, the superconducting proximity is accounted for by a tunneling self-energy $\hat{\Sigma}(\epsilon)$, 
leading to an effective Hamiltonian of the 3D TI in the form
\begin{eqnarray}
&
\hat{H}^{eff}_{\bm p} = \hat{H}_{\bm p} + \hat{\Sigma}(\epsilon),
\qquad 
\hat{H}_{\bm p}=
\Biggl[
\begin{array}{cc}
h_{\bm p} &  0  \\
0 &  -h^*_{-\bm p}
\end{array}
\Biggr]
&
\label{H_def}\\ 
&
\hat{\Sigma}(\epsilon) =
\left[
\begin{array}{cc}
-i\Gamma(\epsilon)\sigma_0   &  \Delta(\epsilon) i\sigma_y {\rm e}^{i\phi}  \\
-\Delta(\epsilon) i\sigma_y {\rm e}^{-i\phi}   &  -i\Gamma(\epsilon) \sigma_0  
\end{array}
\right],
&
\label{Sigma}
\end{eqnarray}
\begin{eqnarray}
&&
\Delta (\epsilon) = i \Gamma_0 f_{_S}( \epsilon ) =  i \Gamma_0 \frac{\Delta_{_S}}{\sqrt{ \epsilon^2 - \Delta^2_{_S} }}, 
\label{Delta}\\
&&
\Gamma(\epsilon) = \Gamma_0 g_{_S}( \epsilon ) = \Gamma_0 \frac{\epsilon}{\sqrt{ \epsilon^2 - \Delta^2_{_S} }},
\,\,\, 
\Gamma_0 =\pi t^2 N_{_S}. \,\,\,
\label{Gamma}
\end{eqnarray}
In Eq. (\ref{H_def}) $h_{\bm p} = v {\bm \sigma} \cdot {\bm p} - \mu$ is the Hamiltonian of the surface state in the absence of tunneling 
(${\bm p}=[p_x,p_y,0]$ is the carrier momentum on the surface, $(\hbar/2){\bm \sigma}$ is the spin vector expressed in terms of Pauli matrices 
$\sigma_x,\sigma_y$, and $\sigma_z$ ($\sigma_0$ is the unit matrix), and $v$ and $\mu$ are the surface Fermi velocity and energy, respectively). The matrix $\hat{H}_{\bm p}$ 
is the representation of the surface-state Hamiltonian in the four-component Nambu basis consisting of the wave functions 
$\psi_{ \uparrow, \bm{p}, \epsilon }, \psi_{ \downarrow, \bm{p}, \epsilon },\psi^*_{ \uparrow, -\bm{p}, -\epsilon }$ and $\psi^*_{ \downarrow, -\bm{p}, -\epsilon }$, where
the star denotes complex conjugation. The tunneling self-energy $\hat{\Sigma}(\epsilon)$ (\ref{Sigma}) has off-diagonal entries describing  
the induced singlet superconducting pairing with strength $\Delta(\epsilon)$ and phase $\phi$. The diagonal entries of $\hat{\Sigma}(\epsilon)$ 
account for the shift $i\Gamma(\epsilon)$ of quasiparticle energies due to the tunneling. Both $\Delta(\epsilon)$ and $\Gamma(\epsilon)$ are energy dependent,
as they are related to the condensate and quasiparticle (momentum-integrated) Green's functions of the S, 
$f_{_S}( \epsilon ) = \Delta_{_S} /\sqrt{ \epsilon^2 - \Delta^2_{_S} }$ and $g_{_S}( \epsilon ) = \epsilon/\sqrt{ \epsilon^2 - \Delta^2_{_S} }$, 
where $\Delta_{_S}$ is the energy gap in the S. 

The characteristic energy scale of $\hat{\Sigma}(\epsilon)$ is $\Gamma_0$, which is determined by the tunneling coupling strength $t$ 
and the normal-state density of states, $N_{_S}$, in the S [see Eq. (\ref{Gamma})]. At high energies $|\epsilon| \gg \Delta_{_S}$, the induced pairing
is suppressed: $\Delta(\epsilon) \sim \Gamma_0\, \Delta_{_S}/|\epsilon|$, whereas $\Gamma(\epsilon) \approx \Gamma_0$, 
yielding the quasiparticle escape rate into the S. In the zero-energy limit $|\epsilon| \to 0$, $\Gamma(\epsilon)$ vanishes, and 
only the off-diagonal entries with constant $\Delta(\epsilon) \approx \Gamma_0$ remain. In this limit the effective Hamiltonian (\ref{H_def}) 
coincides with that of Ref. \onlinecite{Fu08}. In our calculations we will use the more general Hamiltonian (\ref{H_def}) given explicitly by
\begin{equation}
\hat{H}^{eff}_{\bm p} =
\left[
\begin{array}{cc}
 v {\bm \sigma} \cdot {\bm p} - \mu -i\Gamma(\epsilon)   &  \Delta(\epsilon) i\sigma_y {\rm e}^{i\phi}  \\
-\Delta(\epsilon) i\sigma_y {\rm e}^{-i\phi}   &   v {\bm \sigma}^* \cdot {\bm p} + \mu -i\Gamma(\epsilon)   
\end{array}
\right].
\label{H_eff}
\end{equation}

\subsection{Helical quasiparticles, induced gap, and mixed $s$- and $p$-wave pair correlations}
\label{Bogolubov}

The main focus of this article is on the superconducting properties of inhomogeneous systems such as described in Sec.~\ref{ABS}. 
However, before going to that more complex problem, it is necessary to briefly discuss the homogeneous superconducting state described  
by Hamiltonian (\ref{H_eff}). Its main feature is the induced $p$-wave correlations~\cite{Fu08} emerging from an interplay of the singlet $s$-wave pairing 
$\propto i\sigma_y$ and carrier spin helicity ${\bm \sigma} \cdot \hat{\bm p}$, where $\hat{\bm p}$ denotes the unit vector in the momentum direction.   
The precise classification of the pairing symmetry of the induced order parameter as well as the character of the single-particle excitations can be obtained 
from the Green's function $\hat{G}_{\bm p}$ defined by the equation 
\begin{equation}
[\epsilon - \hat{H}^{eff}_{\bm p}]\hat{G}_{\bm p}= \hat{I},
\label{Eq_G}
\end{equation}
with $\hat{I}$ being the $4\times 4$ unit matrix. We will consider $\hat{G}_{\bm p}$ near the Fermi level positioned in the conduction band at high energies
\begin{equation}
\mu \gg \Delta(\epsilon). 
\label{mu}
\end{equation}
Inverting the equation for $\hat{G}_{\bm p}$ we find the matrix elements of the Green's function as follows 
\begin{eqnarray}
\hat{G}_{\bm p}= \left[
\begin{array}{cc}
G_{ 11,{\bm p} }  &  G_{ 12,{\bm p} }  \\
G_{ 21,{\bm p} }  &  G_{ 22,{\bm p} }
\end{array}
\right],
\label{G}
\end{eqnarray}
where the indices $1$ and $2$ refer to the Nambu space, and the corresponding entries are $2\times 2$ matrices in spin space given by~\cite{GT13b}  
\begin{equation}
G_{ 11,{\bm p} } =\frac{1}{2}
\frac{  (\sigma_0 + {\bm \sigma}\cdot \hat{\bm p} )  \, [ {\cal E}(\epsilon) + v (p-p_F) ] } 
     {  {\cal E}^2(\epsilon) - v^2(p-p_F)^2 - \Delta^2(\epsilon) },
\label{G11}
\end{equation}
\begin{equation}
G_{ 22,{\bm p} } = \frac{1}{2}
\frac{   -i\sigma_y (\sigma_0 + {\bm \sigma} \cdot \hat{\bm p} )i\sigma_y \, [ {\cal E}(\epsilon) - v (p-p_F) ] } 
     {  {\cal E}^2(\epsilon) - v^2(p-p_F)^2 - \Delta^2(\epsilon) },
\label{G22}
\end{equation}
\begin{equation}
G_{ 12,{\bm p} } = \frac{1}{2}
\frac{   (\sigma_0 + {\bm \sigma} \cdot \hat{\bm p} )i\sigma_y  \, \Delta(\epsilon) \, {\rm e}^{i\phi} } 
     {  {\cal E}^2(\epsilon) - v^2(p-p_F)^2 - \Delta^2(\epsilon) },
\label{G12}
\end{equation}
\begin{equation}
G_{ 21,{\bm p} } = \frac{1}{2}
\frac{  -i\sigma_y (\sigma_0 + {\bm \sigma} \cdot \hat{\bm p} )\, \Delta(\epsilon) \, {\rm e}^{-i\phi} } 
     {  {\cal E}^2(\epsilon) - v^2(p-p_F)^2 - \Delta^2(\epsilon) },
\label{G21}
\end{equation}
where 
\begin{eqnarray}
{\cal E}(\epsilon) = \epsilon + i\Gamma(\epsilon), \quad \hat{\bm p} = {\bm p}/p_F, \quad 
\sigma_0 = \left(
  \begin{matrix}
    1 & 0 \\
    0 & 1 
  \end{matrix}
\right).
\label{E_eps}
\end{eqnarray}
Equation (\ref{G11}) for $G_{ 11|{\bm p} }$ reveals properties of the single-particle excitations in the superconducting TI. 
Just like in the normal TI (cf. Ref. \onlinecite{GT11}) they are eigenstates of the helicity ${\bm \sigma}\cdot \hat{\bm p}$, 
but their spectrum is modified due to the superconducting proximity effect and is obtained by solving the equation  
\begin{equation}
{\cal E}^2(\epsilon)  - v^2(p-p_F)^2 - \Delta^2(\epsilon) = 0.
 \label{Eq_Spectrum}
\end{equation}
For $p \approx p_F$ the solution to Eq. (\ref{Eq_Spectrum}) is
\begin{equation}
\epsilon_{\bm p} \approx \pm \sqrt{ v^2(p-p_F)^2 + \varepsilon^2_{\rm g} },
\label{Spectrum}
\end{equation}
where $\varepsilon_{\rm g}$ is the induced superconducting gap given by~\cite{GT13b}   
\begin{eqnarray}
\varepsilon_{\rm g} \approx \Gamma_0 \left(1 - \gamma + \frac{3}{2} \gamma^2\right),
\quad 
\gamma = \frac{\Gamma}{\Delta_{_S}} \ll 1. 
\label{Eg}
\end{eqnarray}
\begin{figure}[b]%
\begin{center}  
\includegraphics[width=0.6\linewidth]{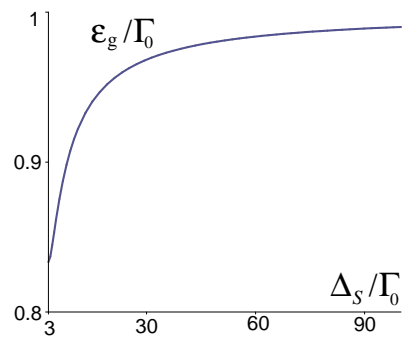}%
\end{center}
\caption{%
Induced gap $\varepsilon_{\rm g}$ (\ref{Eg}) versus superconducting gap $\Delta_{_S}$. 
}
\label{Gap}
\end{figure}
We introduced the dimensionless parameter $\gamma$ assumed small throughout.
According to Eq. (\ref{Eg}), $\varepsilon_{\rm g}$ is reduced as the gap in the superconductor, $\Delta_{_S}$, becomes smaller. 
This tendency is shown in Fig. \ref{Gap}. In Sec. \ref{Current} we will encounter a similar dependence of the critical current in the S/TI/S junctions.  
It is also worth noting that the energy $\varepsilon_{\rm g}$ can be extracted from the temperature dependence of the critical current 
in short proximity-effect junctions.~\cite{Rohlfing09}

The Green's function $G_{ 21|{\bm p} }$ (\ref{G21}) (as well as $G_{ 12|{\bm p} }$) 
describes the induced superconducting condensate. The symmetry of the induced pair correlations is identified by the spin structure of Eq. (\ref{G21}):
\begin{equation}
G_{ 21|{\bm p} } \propto (i\sigma_y + i\sigma_y {\bm \sigma} \cdot \hat{\bm p}) \Delta(\epsilon).
\label{Pair}
\end{equation}
Here the first term is a singlet $s$-wave component, whereas the second term is a triplet $p$-wave pairing in which  
the unit momentum vector, $\hat{\bm p}$, can be identified as the Balian-Werthamer $d$-vector 
(for a more detailed classification of the pairing of spinfull electrons, see, e.g., Ref. \onlinecite{Leggett75}).
Equation (\ref{Pair}) implies the usual basis of the up- and down-spin states and 
agrees with the results of Ref. \onlinecite{Stanescu10} for a large Fermi energy. 
The origin of the mixed $s$- and $p$-wave superconducting correlations is the broken spin-rotation symmetry 
due to the helicity of the surface states. The situation reminds to some extent the mixed singlet-triplet intrinsic superconductivity predicted 
for systems without inversion symmetry. \cite{Gorkov01,Santos10} The induced $p$-wave component inherits the spin-momentum locking 
${\bm \sigma} \cdot \hat{\bm p}$ of the normal-state carriers. Since it is an odd function of the momentum direction $\hat{\bm p}$, 
the $p$-wave component is suppressed in dirty TIs~\cite{GT13b} in which the elastic mean-free path $\ell$ is much smaller than the superconducting coherence length 
$\xi$. In this paper we consider the opposite case of a clean TI defined by the condition:
\begin{equation}
 \ell \gg \xi, \quad \xi = \hbar v/\sqrt{ \Delta^2(\epsilon) - {\cal E}^2(\epsilon) }.
\label{clean}
\end{equation}
In the next sections we discuss the link between the mixed $s$- and $p$-wave pairing (\ref{Pair}) 
and Andreev bound states (ABSs) in a ballistic S/TI/S Josephson junction.

\section{Helical Andreev bound states}
\label{ABS}

Let us consider a junction between two superconducting TI terminals (see Fig. \ref{STIS}). 
The junction length defined as distance between the terminals is assumed much smaller than the superconducting coherence length $\xi$ [see Eq. (\ref{clean})].   
Concretely, for Nb/TI contacts with typical parameters 
$\Delta_{_S} \approx 1$ meV, $\Gamma_0 \approx 0.2$ meV, and $\hbar v \approx 250 \div 350$ meV nm (see e.g. Ref.\onlinecite{Maier12}),  
the junction length should be smaller than $1.25 \div 1.75$ $\mu$m, which is easily realizable experimentally. 
In this limit the junction can be described by an equation  
\begin{eqnarray}
[H^{eff}( {\bm r} ) +  \tau_z\sigma_{_0} U(x) ] \Psi({\bm r}) = \epsilon \,\Psi({\bm r}), 
\label{Eq_Psi}
\end{eqnarray}
for the Nambu spinor $\Psi({\bm r})$ in position representation.
Explicitly, Eq. (\ref{Eq_Psi}) can be written as
\begin{widetext}
\begin{equation}
\left[
\begin{array}{cccc}
U(x) - \mu - {\cal E}(\epsilon)    & \hbar v(-i\partial_x - \partial_y)       & 0                                   & \Delta(\epsilon) {\rm e}^{ i\phi_{_{L,R} } } \\
\hbar v(-i\partial_x + \partial_y) & U(x) - \mu - {\cal E}(\epsilon)          & -\Delta(\epsilon){\rm e}^{i\phi_{_{L,R} } }    & 0\\
0                                  & -\Delta(\epsilon){\rm e}^{-i\phi_{_{L,R} } }        & \mu - U(x) -  {\cal E}(\epsilon)    & \hbar v(-i\partial_x + \partial_y) \\ 
\Delta(\epsilon){\rm e}^{-i\phi_{_{L,R} } }   & 0                                        & \hbar v(-i\partial_x - \partial_y)  & \mu - U(x) -  {\cal E}(\epsilon)
\end{array}
\right]
\left[
\begin{array}{c}
\psi_{ \uparrow, \epsilon }({\bm r}) \\ \psi_{ \downarrow,\epsilon }({\bm r}) \\ \psi^*_{ \uparrow, -\epsilon}({\bm r}) \\ \psi^*_{ \downarrow, -\epsilon }({\bm r})
\end{array}
\right] =0 , \quad \phi_{_L}=0 \quad \phi_{_R}=\phi,
\label{Eq_Psi1}
\end{equation}
\end{widetext}
where we introduce different phases $\phi_{L,R}$
in the left (L, $x<0$) and right (R, $x>0$) terminals. 
In addition to that we introduce a potential $U(x) = U \delta(x)$ in the middle of the junction 
(the Pauli matrix $\tau_z$ in Eq. (\ref{Eq_Psi}) acts in the Nambu space). 
In practice, the potential barrier can be induced by a top-gate electrode as sketched in Fig. \ref{STIS}.
The precise form of the scattering potential is not of crucial importance here 
because we will focus on the effects arising from Klein tunneling that takes place independently of the barrier shape.  
We note that the related earlier studies of Refs. \onlinecite{Fu08,Olund12} and \onlinecite{Yamakage12} assumed a scattering-free weak link. 
The influence of the normal scattering on the ABSs and the resulting superconducting Klein tunneling  
was first pointed out in Ref. \onlinecite{GT13}.

The ABS solutions of Eq. (\ref{Eq_Psi}) occur at energies $\epsilon$ smaller than the induced gap and decay exponentially for $x \to \pm \infty$.
We seek such solutions as the sum of independent modes with different $k_y$ wave numbers:  
\begin{equation}
\Psi({\bm r}) = \sum_{k_y} \Psi_{k_y}(x) \frac{ {\rm e}^{ik_y y} }{ \sqrt{W} }.
\label{Sum}
\end{equation}
where $\Psi_{k_y}(x)$ are evanescent functions for both $x > 0$ and $x<0$. 
For a large Fermi energy (\ref{mu}) the spinors $\Psi_{k_y}(x)$ are readily obtained from Eq. (\ref{Eq_Psi}) as follows 
\begin{eqnarray}
 \Psi_{k_y} (x>0) &=&  
\left[
\begin{array}{c}
 1 \\ {\rm e}^{i\theta} \\ -{\rm e}^{i\theta} a^+_R \\  a^+_R 
\end{array}
\right]
C^>_R 
\,
{\rm e}^{ ix k_F\cos\theta - x/\xi }
\label{Psi_R}\\
& + &
\left[
\begin{array}{c}
 1 \\ -{\rm e}^{-i\theta} \\ {\rm e}^{-i\theta} a^-_R \\  a^-_R 
\end{array}
\right]
C^<_R 
\,
{\rm e}^{ -ix k_F\cos\theta - x/\xi },
\nonumber
\end{eqnarray}
\begin{eqnarray}
 \Psi_{k_y} (x<0) &=&  
\left[
\begin{array}{c}
 1 \\ {\rm e}^{i\theta } \\ -{\rm e}^{i\theta } a^-_L \\  a^-_L 
\end{array}
\right]
C^>_L 
\,
{\rm e}^{ ix k_F\cos\theta + x/\xi }
\label{Psi_L}\\
& + &
\left[
\begin{array}{c}
 1 \\ -{\rm e}^{-i\theta } \\ {\rm e}^{-i\theta } a^+_L \\  a^+_L 
\end{array}
\right]
C^<_L 
\,
{\rm e}^{ -ix k_F\cos\theta + x/\xi },
\nonumber
\end{eqnarray}
where $C^{>,<}_{R,L}$ are constants, the length-scale $\xi$ is defined in Eq. (\ref{clean}) and other parameters and notations are explained below: 
\begin{eqnarray} 
&&
a^\pm_{R,L}(\epsilon) = 
\frac{ \Delta(\epsilon){\rm e}^{ -i \phi_{_{R,L} }  } }
     { {\cal E}(\epsilon) \pm i\sqrt{  \Delta^2 (\epsilon) - {\cal E}^2(\epsilon)  } },
\label{a}\\
&&
\cos\theta = \sqrt{ 1 - (k_y/k_F)^2}
\label{theta}
\end{eqnarray}
Here $a^\pm_{R,L}(\epsilon)$ are the Andreev reflection amplitudes at the ends of the junction, 
the angle $\theta$ indicates the particle propagation direction (spanning the range from $-\pi/2$ to $\pi/2$ around the $x$-axis) 
for given open channel with $k_y$ smaller than the Fermi wave number $k_F$.
The matching condition for spinors (\ref{Psi_R}) and (\ref{Psi_L}) is obtained by integrating Eq. (\ref{Eq_Psi}) around $x=0$, which yields 
\begin{equation} 
\Psi_{k_y}(-0) = ( 1 + iZ\tau_z\sigma_x )\Psi_{k_y}(+0), \quad Z = U/\hbar v,
 \label{Matching}
\end{equation}
where the dimensionless parameter $Z$ characterizes the barrier strength. 
Inserting (\ref{Psi_R}) and (\ref{Psi_L}) into (\ref{Matching}) we arrive at four linear algebraic equations for the four constants  
$C^>_R$,  $C^<_R$, $C^>_L$, and $C^<_L$. Equating the determinant of the system to zero, 
we then obtain after some algebra the following equation for the ABSs:
\begin{eqnarray}
& [a^+_R(\epsilon) - a^-_R(\epsilon)][a^+_L(\epsilon) - a^-_L(\epsilon)] = &
\nonumber\\
& T_\theta\,[a^+_R(\epsilon) - a^+_L(\epsilon)][a^-_R(\epsilon) - a^-_L(\epsilon)], &
\label{Eq_a}
\end{eqnarray}
where $T(\theta)$ is the normal-state junction transparency:
\begin{equation}
T_\theta = \frac{ \cos^2(\theta) } 
           { 1 - \sin^2(\theta)/(1+Z^2) }. 
\label{T}    
\end{equation}
Using Eq. (\ref{a}) for the Andreev amplitudes, we obtain from (\ref{Eq_a}) an equation for $\epsilon$:~\cite{Link_GT13}

\begin{equation}
\left(  \frac{\epsilon}{ \Delta_{_S} } + \frac{\epsilon}{ \Gamma_0 } \sqrt{  1 - \frac{\epsilon^2}{ \Delta^2_{_S} }  } \right)^2 
= 1 - T_\theta \sin^2 \frac{\phi}{2}.
\label{Eq_E}
\end{equation}
Its solutions depend not only on the phase difference $\phi$, but also on the relative energy $\Gamma_0/\Delta_{_S}$ 
characterizing the S/TI interface. The latter dependence was not accounted for in the previous calculations of the ABSs in S/TI/S junctions. 

The solutions of Eq. (\ref{Eq_E}) come in pairs $\epsilon^{\pm} = \pm |\epsilon|$. 
In order to find $|\epsilon|$ it is convenient to recast Eq. (\ref{Eq_E}) as follows:
\begin{equation}
\left| \frac{\epsilon}{\Delta_{_S}} \right|^2 \left( 1-  \left| \frac{\epsilon}{\Delta_{_S}} \right|^2 \right) =
\gamma^2 \, 
\left( 
\sqrt{
1 - T_\theta \sin^2 \frac{\phi}{2}
} 
-
\left|
\frac{\epsilon}{\Delta_{_S}}
\right|
\right)^2, 
\label{Eq_E1}
\end{equation}
where $\gamma$ is the small parameter defined in Eq. (\ref{Eg}). 
We seek the solution to Eq. (\ref{Eq_E1}) in the form of the expansion in powers of $\gamma$:
\begin{equation}
\frac{|\epsilon|}{\Delta_{_S}} = \gamma f_1(\phi) + \gamma^2 f_2(\phi) + \gamma^3 f_3(\phi) +... 
\label{Exp_ABS}
\end{equation}
The functions $f_1(\phi), f_2(\phi), f_3(\phi) ...$ are obtained by equating the coefficients at $\gamma, \gamma^2, \gamma^3 ...$ 
on the left- and right-hand sides of Eq. (\ref{Eq_E1}). 
Up to the cubic terms we find 
\begin{eqnarray}
&&
f_1 = -f_2 = \left[ 1 - T_\theta \sin^2 \frac{\phi}{2} \right]^{1/2},
\label{f12}\\
&&
f_3 = \left[ 1 - T_\theta \sin^2 \frac{\phi}{2} \right]^{1/2} + \frac{1}{2} \left[ 1 - T_\theta \sin^2 \frac{\phi}{2} \right]^{3/2}.
\label{f3}
\end{eqnarray}
Within this approximation the ABSs are given by~\cite{Link_GT13}
\begin{eqnarray}
&&
E^\pm_{\theta}(\phi) \approx \pm \Gamma_0 
\left(  (  1 - \gamma + \gamma^2  ) \left[ 1 - T_\theta \sin^2 \frac{\phi}{2} \right]^{1/2} + \right.
\nonumber\\
&&
+
\left.
\frac{\gamma^2}{2} \left[ 1 - T_\theta \sin^2\frac{\phi}{2} \right]^{3/2} \right).
\label{ABSs}
\end{eqnarray}
The ABSs appear exactly below the induced gap given by (\ref{Eg}), i.e. $|E^\pm_{\theta}(\phi)| \leq \varepsilon_{\rm g} = |E^\pm_{\theta}(0)|$.

\begin{figure}[t]%
\begin{center}  
\includegraphics[width=0.7\linewidth]{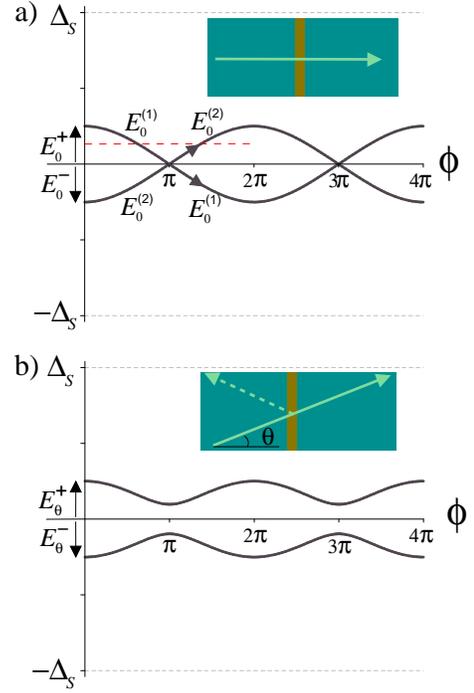}%
\end{center}
\caption{%
Andreev bound states (\ref{ABSs}) versus phase difference $\phi$ for (a) normal incidence at potential barrier with $\theta = 0$ and 
(b) oblique incidence with $\theta = \pi/4$; $\gamma = 0.3$, $Z=0.5$. Arrows indicate the $4\pi$-periodic branches (\ref{ABSs_4pi}). 
Dashed line corresponds to fixed energy (\ref{Energy}).
}
\label{ABSs_fig}
\end{figure}

\subsection{Angle-selective protection of ABS$s$}

Figure \ref{ABSs_fig} shows the ABSs for two distinct cases: normal incidence at the potential barrier with $\theta=0$ (panel a)
and oblique incidence with $\theta\not = 0$ (panel b). The perpendicular propagation is reflectionless, with the transparency $T_0 = 1$ 
for any barrier strength [see Eq.(\ref{T}) and Fig. \ref{STIS}b], which is the familiar Klein tunneling. In this case, 
the ABSs are gapless (see also Fig. \ref{ABSs_fig}a)
\begin{eqnarray}
E^\pm_{0}(\phi) \approx \pm \Gamma_0 
\left[  (  1 - \gamma + \gamma^2  ) \left| \cos\frac{\phi}{2} \right| +
\frac{\gamma^2}{2} \left| \cos\frac{\phi}{2} \right|^{3} \right].
\label{ABSs_0}
\end{eqnarray}
We recall that gapless ABSs also appear in conventional ballistic Josephson junctions.~\cite{Beenakker04} 
Unlike the latter, the solutions (\ref{ABSs_0}) are protected by time-reversal symmetry. 
In order to check this we notice that the $\pm$ states in Eq. (\ref{ABSs_0}) are equivalent to two $4\pi$-periodic branches 
crossing in the middle of the gap:
\begin{eqnarray}
E^{(1,2)}_{0}(\phi) \approx \pm \Gamma_0 
\left[  (  1 - \gamma + \gamma^2  ) \cos\frac{\phi}{2} + \frac{\gamma^2}{2} \cos^3\frac{\phi}{2} \right],
\label{ABSs_4pi}
\end{eqnarray}
where we introduced notation $E^{(1,2)}_{0}(\phi)$ to distinguish the $4\pi$-periodic solutions (\ref{ABSs_4pi}) from (\ref{ABSs_0}). 
Let us now examine their wave functions at the same fixed energy [shown by the dashed line in Fig. \ref{ABSs_fig}(a)],  
which implies
\begin{equation}
E^{(1)}_0(\phi)=E^{(2)}_0(2\pi -\phi)=\epsilon.
\label{Energy}
\end{equation}
The wave functions of these states [see, e.g., Eq. (\ref{Psi_R}) for $\theta =0$] appear to consist only of one-way propagating states:
\begin{eqnarray}
&&
\Psi^{(1)}_0(x) =  
\left[
\begin{array}{c}
 1 \\ 1\\ -a^+_R(\phi)\\ a^+_R(\phi) 
\end{array}
\right]
\frac{  {\rm e}^{ ik_Fx - x/\xi }  }{ 2\sqrt{\xi} }, 
\label{Psi_1}\\
%
%
&&
\Psi^{(2)}_0(x) = 
\left[
\begin{array}{c}
 1 \\ -1\\ a^-_R(2\pi-\phi)\\a^-_R(2\pi-\phi) 
\end{array}
\right]
\frac{  {\rm e}^{ -ik_Fx - x/\xi }  }{ 2\sqrt{\xi} }, 
\label{Psi_2}
\end{eqnarray}
where $1/2\sqrt{ \xi }$ is the normalization factor. Note that for the given energy (\ref{Energy}) the Andreev amplitude $a^-_R$ in $\Psi^{(2)}_0(x)$  
has the phase $2\pi - \phi$. It is now easy to see that the states (\ref{Psi_1}) and (\ref{Psi_2}) 
are connected by the time-reversal operation:
\begin{equation}
\Psi^{(2)}_0(x) = \tau_0 i\sigma_y \Psi^{(1) *}_0(x).
\label{TR}
\end{equation}
where $\tau_0$ denotes a unit matrix in Nambu space. Moreover, $\Psi^{(1)}_0(x)$ and $\Psi^{(2)}_0(x)$ 
are the eigenstates of the helicity matrix $\tau_z\sigma_x$: 
\begin{equation}
\tau_z\sigma_x \Psi^{(1,2)}_0(x) = \pm \Psi^{(1,2)}_0(x).
\label{Helicity}
\end{equation}
This shows that the spin of these states is tied to the momentum direction. As a consequence of Eqs. (\ref{TR}) and (\ref{Helicity}), 
the ABSs (\ref{Psi_1}) and (\ref{Psi_2}) are orthogonal to each other and, therefore, 
immune to spin-independent potential scattering.

Clearly, the topological protection does not hold for the ABSs with oblique incidence at the potential barrier since in this 
case the incident and reflected trajectories are not related by the time reversal [see, Fig. \ref{ABSs_fig}(b)]. 
The reflection from the potential barrier generates a minigap $\propto \sqrt{ 1 - T_\theta }$ at $\phi=\pi, 3\pi,...$, 
similar to the ABSs in conventional Josephson junctions.~\cite{Beenakker04}   

Although for $\theta \not =0$ the ABSs are not protected from potential scattering, they still feature the spin-momentum locking 
and, therefore, can be called helical in the same sense as the normal-state 2D surface carriers.  
It is worth noting that despite the dependence on the normal-state quantity $T_\theta$ 
the helical ABSs reflect the nature of the induced superconducting condensate, i.e., its mixed $s$- and $p$-wave character.
In Eq. (\ref{Eq_Psi1}) the $s$- and $p$-wave correlations are explicitly accounted for by the couplings between the opposite-spin  
(e.g., $\psi_{\uparrow, \epsilon}$ and $\psi^*_{\downarrow, -\epsilon}$) and the same-spin 
(e.g., $\psi_{\uparrow, \epsilon}$ and $\psi^*_{\uparrow, -\epsilon}$) particle and hole components, respectively. 
Alternatively, one can examine the spin structure of the Green's function of the ABSs. It can be obtained by means of   
the Hilbert-Schmidt expansion in terms of the eigen-spinors and energies of the ABSs. 
Since for our purpose the full expansion is not needed, we focus on the contributions of the normally propagating states (\ref{Psi_1}) and (\ref{Psi_2}).
For $\epsilon \to E^{(1)}_0(\phi)$, the contribution of the state (\ref{Psi_1}) is constructed by making a direct product ($\otimes$) of spinor $\Psi^{(1)}_0(x)$ and its conjugate 
$\tilde{\Psi}^{(1)}_0(x^\prime)=\left[  1 , 1 , -a^{+*}_R , a^{+*}_R \right]\, {\rm e}^{-ik_Fx^\prime - x^\prime/\xi}/(2\sqrt{\xi})$ as follows
\begin{widetext}
\begin{eqnarray}
G^{(1)}_0(x,x^\prime) \approx \frac{ \Psi^{(1)}_0(x) \otimes \tilde{\Psi}^{(1)}_0(x^\prime) }{ \epsilon - E^{(1)}_0(\phi) }
&=&
\left[
\begin{array}{cccc}
1 & 1 & -a^{+*}_R & a^{+*}_R \\ 
1 & 1 & -a^{+*}_R & a^{+*}_R \\
-a^+_R & -a^+_R & 1 & -1 \\ 
a^+_R & a^+_R & -1 & 1
\end{array}
\right]_{\epsilon =E^{(1)}_0(\phi) }
\times
\frac{ {\rm e}^{ik_F(x-x^\prime) - (x+x^\prime)/\xi } }{4\xi\, [\epsilon - E^{(1)}_0(\phi)]}
\label{G_1}\\
&=&
\left[
\begin{array}{cc}
\sigma_0 + \sigma_x  & (\sigma_0 +\sigma_x)i\sigma_y a^{+*}_R\\ 
-i\sigma_y (\sigma_0 +\sigma_x)a^{+}_R & \sigma_0 - \sigma_x   
\end{array}
\right]_{\epsilon =E^{(1)}_0(\phi) }
\times
\frac{ {\rm e}^{ik_F(x-x^\prime) - (x+x^\prime)/\xi} }{ 4\xi\, [\epsilon - E^{(1)}_0(\phi)]}.
\label{G_1_M}
\end{eqnarray}
\end{widetext}
Clearly, the spin structure of the diagonal and off-diagonal matrix elements of Eq. (\ref{G_1_M}) is identical to that of 
the Green's functions in Eqs. (\ref{G11}) -- (\ref{G21}) for $\hat{\bm p} || x$. 
This proves that ABS (\ref{Psi_1}) is of the mixed  $s-$ and $p$-wave type.
Similar calculation for the counter-propagating state (\ref{Psi_2}) yields
\begin{widetext}
\begin{eqnarray}
&&
G^{(2)}_0(x,x^\prime) \approx \frac{ \Psi^{(2)}_0(x) \otimes \tilde{\Psi}^{(2)}_0(x^\prime) }{ \epsilon - E^{(2)}_0(\phi) }
=
\left[
\begin{array}{cc}
\sigma_0 - \sigma_x  & (\sigma_0 - \sigma_x)i\sigma_y a^{+}_R\\ 
-i\sigma_y (\sigma_0 -\sigma_x)a^{+*}_R & \sigma_0 + \sigma_x   
\end{array}
\right]_{\epsilon =E^{(2)}_0(\phi) }
\times
\frac{ {\rm e}^{-ik_F(x-x^\prime) - (x+x^\prime)/\xi } }{4\xi\, [\epsilon - E^{(2)}_0(\phi)]}.
\label{G_2_M}
\end{eqnarray}
\end{widetext}
The spin matrices in Eqs. (\ref{G_1_M}) and (\ref{G_2_M}) are related to each other by the time-reversal operation.

\section{Current-phase relation $I(\phi)$, critical current I$_c$ and I$_c$R$_N$ product}
\label{Current}

\begin{figure}[b]%
\begin{center}  
\includegraphics[width=0.80\linewidth]{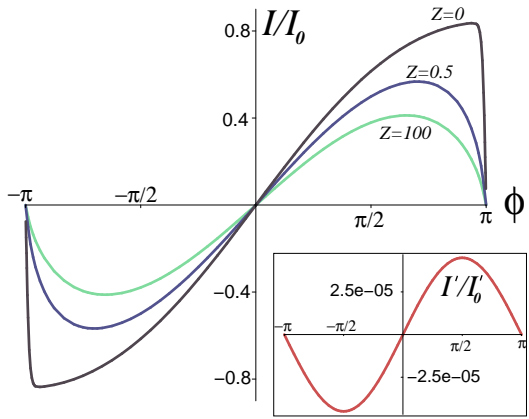}%
\end{center}
\caption{%
Current-phase relation $I(\phi)$ (\ref{I}) for different barrier strengths $Z$. 
Inset: current-phase relation $I^\prime(\phi)$ (\ref{I'}) for a reference conventional S/N/S junction 
for strong barrier with $Z=100$. In both cases $\gamma=0.2$ and $\Xi = 0.01 \Gamma_0$.   
}
\label{CPR}
\end{figure}

In the short junction case the Josephson current is carried mostly by the ABSs and can be calculated in equilibrium 
using the formula:~\cite{Beenakker04}
\begin{equation}
 I(\phi) = - \frac{eN}{\hbar} 
\int\limits_0^{\pi/2} d\theta \cos\theta \, \frac{ \partial E^+_\theta (\phi) }{ \partial \phi }\, \tanh \frac{ E^+_\theta (\phi) }{2 \Xi}, 
\label{I_def}
\end{equation}
where $N = k_F W/\pi$ is the number of the open channels and $\Xi$ is the temperature in energy units. 
For the ABSs in Eq. (\ref{ABSs}) the current (\ref{I_def}) is
\begin{widetext}
\begin{equation}
\frac{I(\phi)}{I_0} = \frac{\sin\phi}{2} 
\int\limits_0^{\pi/2} d\theta  \cos\theta   
\frac{ T_\theta [1 - \gamma + \frac{5}{2}\gamma^2 - \frac{3}{2}\gamma^2 T_\theta \sin^2 \frac{\phi}{2} ] }{ \left[ 1 - T_\theta \sin^2 \frac{\phi}{2} \right]^{1/2} }
\,
\tanh 
\frac{ 
(  1 - \gamma + \gamma^2  ) \left[ 1 - T_\theta \sin^2 \frac{\phi}{2} \right]^{1/2} 
+
\frac{1}{2} \gamma^2 \left[ 1 - T_\theta \sin^2\frac{\phi}{2} \right]^{3/2}  
}
{ 2\Xi/\Gamma_0 }
\label{I}
\end{equation}
where $I_0 = eN\Gamma_0/2\hbar$. It is instructive to also consider a reference S/N/S junction of the same geometry, 
where N is a conventional 2D system with a spin-degenerate parabolic dispersion. In this system the proximity effect 
is described by the same self-energy as in Eq. (\ref{Sigma}). For the Josephson current in the reference S/N/S system we introduce notation 
$I^\prime(\phi)$. The latter is given by the equation 
\begin{equation}
\frac{I^\prime(\phi)}{I^\prime_0} = \frac{\sin\phi}{2}    
\frac{ T^\prime [1 - \gamma + \frac{5}{2}\gamma^2 - \frac{3}{2}\gamma^2 T^\prime \sin^2 \frac{\phi}{2} ] }
{ \left[ 1 - T^\prime \sin^2 \frac{\phi}{2} \right]^{1/2} }
\,
\tanh 
\frac{ 
(  1 - \gamma + \gamma^2  ) \left[ 1 - T^\prime \sin^2 \frac{\phi}{2} \right]^{1/2} 
+
\frac{1}{2} \gamma^2 \left[ 1 - T^\prime \sin^2\frac{\phi}{2} \right]^{3/2}  
}
{ 2\Xi/\Gamma_0 },
\label{I'}
\end{equation}
where $I^\prime_0 = 2 I_0$ includes the factor of two due to the spin degeneracy and 
$T^\prime = 1/(1+Z^2)$ is the normal-state transparency of the N region with the same (delta-like) potential in the middle. 
For the non-helical carriers in the N neither $T^\prime$ nor $I^\prime(\phi)$ is angle dependent.     
\end{widetext}
The equilibrium current-phase relation (\ref{I}) is $2\pi$-periodic. 
It is plotted for different barrier strengths $Z$ in Fig. \ref{CPR}. 
Interestingly $I(\phi)$ remains non-sinusoidal even for very large $Z$. 
In contrast, for the conventional current (\ref{I'}) for $Z \gg 1$ we get $I^\prime(\phi) \propto \sin\phi$, shown for comparison in the inset of Fig. \ref{CPR}. 
This difference reflects the helical character of the ABSs (\ref{ABSs}) most pronounced in the gapless case $\theta=0$.  

\begin{figure}[b]%
\begin{center}  
\includegraphics[width=0.80\linewidth]{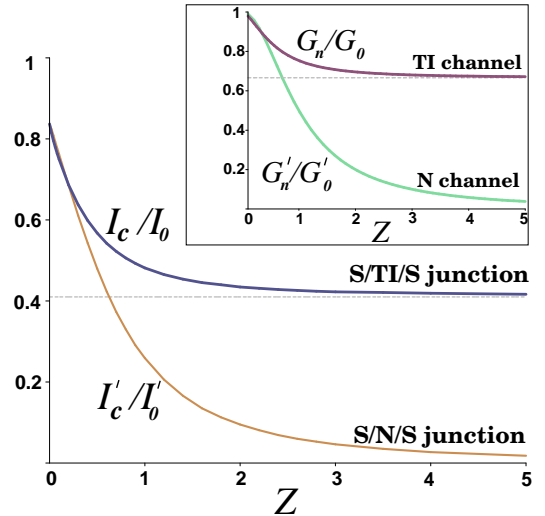}%
\end{center}
\caption{%
Normalized critical current versus barrier strength $Z$ for S/TI/S and conventional S/N/S junctions; 
$\gamma=0.2$ and $\Xi = 0.01 \Gamma_0$.
Inset: Normal-state conductance versus barrier strength $Z$ for TI and N channels (see also the text).
}
\label{Ic}
\end{figure}

Next we analyze numerically the critical current $I_c$ defined as the maximum of $I(\phi)$ (\ref{I}) at fixed parameters $Z$ and $\gamma$.  
Figure \ref{Ic} shows the normalized $I_c$ as function of the barrier strength $Z$. 
Initially decreasing with $Z$, the critical current saturates at a constant value for $Z > 3$.   
This is in sharp constrast with the conventional S/N/S junction for which $I^\prime_c(Z)$ is strongly suppressed 
for $Z > 3$. 

The large supercurrent despite the presence of the strong barrier is a manifestation of the helical ABSs (\ref{ABSs}) 
that are able to transport Cooper pairs without reflection for $\theta =0$ or with weak scattering for $\theta \not =0$ across the junction.  
This can be viewed as a superconducting analog of the normal-state transport dominated by the Klein tunneling. 
For comparison, the inset of Fig. \ref{Ic} shows the normal-state (Sharvin) conductance of the TI channel:
\begin{equation}
G_n = G_0 \int\limits_0^{\pi/2} d\theta \cos\theta \, T_\theta, \quad G_0 = \frac{e^2 N}{h},
\label{G_n}
\end{equation}
along with the conductance of the conventional N channel $G^\prime_n = G^\prime_0 T^\prime$, with $G^\prime_0 = 2G_0$.  

\begin{figure}[t]%
\begin{center}  
\includegraphics[width=0.70\linewidth]{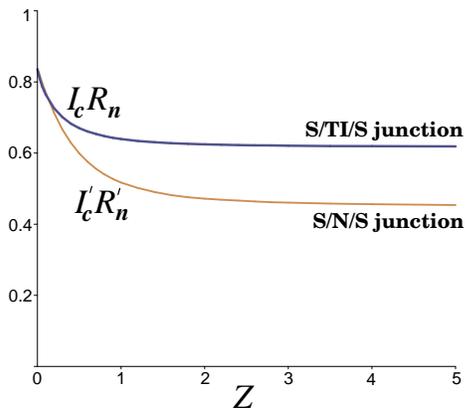}%
\end{center}
\caption{%
$I_cR_n$ product in units of $\pi \Gamma_0/e$ versus barrier strength $Z$ for S/TI/S and conventional S/N/S junctions;
$\gamma=0.2$ and $\Xi = 0.01 \Gamma_0$.
}
\label{JcRn_Z}
\end{figure}

Figure \ref{JcRn_Z} shows another experimentally relevant characteristic, 
the product of the critical current $I_c$ and the normal-state resistance $R_n = 1/G_n$ (\ref{G_n}) normalized by $\pi \Gamma_0/\pi$. 
In accord with the above discussion, the $I_cR_n$ product is larger for the S/TI/S junction compared with the conventional S/N/S system. 

\begin{figure}[t]%
\begin{center}  
\includegraphics[width=0.70\linewidth]{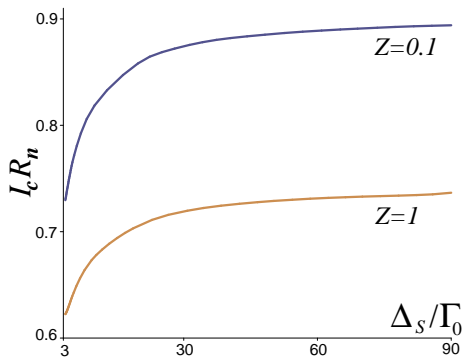}%
\end{center}
\caption{%
$I_cR_n$ product in units of $\pi \Gamma_0/e$ versus normalized superconducting gap $\Delta_{_S}$; 
$\Xi = 0.01 \Gamma_0$. 
}
\label{JcRn_D}
\end{figure}

Finally, in Fig. \ref{JcRn_D} we plot the dependence of the normalized $I_cR_n$ product on the superconducting gap $\Delta_{_S}$.    
The limit $\Delta_{_S}/\Gamma_0 \to \infty$ corresponds to a frequently used simplified model of the proximity effect in 
which the induced pairing potential is approximated by an energy-independent constant $\Delta = \epsilon_{\rm g} =\Gamma_0$. 
In this limit and for small $Z$ we recover the known result
\begin{equation}
I_cR_n \approx e\Gamma_0/\pi   
\label{IcRn}
\end{equation}
However, as shown in Fig. \ref{JcRn_D}, for any finite $\Delta_{_S}/\Gamma_0$ the $I_cR_n$ product is smaller than (\ref{IcRn}). 
Thus, the dependence on the microscopic parameters $\Delta_{_S}$ and $\Gamma_0$ may be one of the factors contributing to 
lower values of the $I_cR_n$ product in proximity-effect Josephson junctions.

\section{Summary}
\label{Summary}

Based on a microscopic model for the superconducting proximity effect, 
we studied the Josephson effect in a short ballistic junction made on the surface of a 3D TI. 
The induced superconductivity in the TI is the mixture of singlet $s$-wave and triplet $p$-wave pair correlations.
Furthermore, the presence of the $p$-wave correlations is encoded in helical Andreev bound states characterized by the spin-momentum locking. 
We showed that the supercurrent carried by helical Andreev bound states decreases in the presence of the potential barrier in the junction,
but cannot be completely suppressed due to the Klein tunneling of Cooper pairs through the barrier. 
Such superconducting Klein tunneling reveals itself in the non-sinusoidal current-phase relation and 
in the saturation of the critical current at high barrier strengths. 
This suggests that the p-wave superconductivity should be detectable in gated TI Josephson junctions.
For that the gate-induced potential must be higher than the Fermi energy of the surface state. 
In this limit we also do not expect the shape of the induced potential barrier to play an essential role.

\acknowledgments
We thank L. W. Molenkamp, B. Trauzettel, P. Recher, J. B. Oostinga and P. Virtanen for valuable discussions.   
This work was supported by the German research foundation (DFG), Grants No HA5893/5-2 (within FOR1162) and TK60/1-1.


\begin{thebibliography}{99}

\bibitem{Fu08} 
L. Fu and C. L. Kane, Phys. Rev. Lett. {\bf 100}, 096407 (2008).

\bibitem{Hasan10}
M. Z. Hasan and C. L. Kane, Rev. Mod. Phys. {\bf 82}, 3045 (2010). 

\bibitem{Qi11} 
X.-L. Qi and S.-C. Zhang, Rev. Mod. Phys. {\bf 83}, 1057 (2011).

\bibitem{Tanaka12} 
Y. Tanaka, M. Sato, and N. Nagaosa, J. Phys. Soc. Jpn. {\bf 81}, 011013 (2012). 

\bibitem{Alicea12}
J. Alicea, Rep. Prog. Phys. {\bf 75}, 076501 (2012). 

\bibitem{Beenakker13a}
C. W. J. Beenakker, Annu. Rev. Condens. Matter Phys. {\bf 4}, 113 (2013).

\bibitem{GT13} 
G. Tkachov and E. M. Hankiewicz, Phys. Status Solidi B {\bf 250}, 215 (2013).




\bibitem{D_Zhang11} 
D. Zhang, J. Wang, A. M. DaSilva, J. S. Lee, H. R. Gutierrez, M. H. W. Chan, J. Jain, and N. Samarth,
Phys. Rev. B {\bf 84}, 165120 (2011). 

\bibitem{Koren11} 
G. Koren, T. Kirzhner, E. Lahoud, K. B. Chashka, and A. Kanigel,
Phys. Rev. B {\bf 84}, 224521 (2011). 

\bibitem{Sacepe11} %
B. Sacepe, J. B. Oostinga, J. Li, A. Ubaldini, N.J.G. Couto, E. Giannini, and A. F. Morpurgo, 
Nat. Commun. {\bf 2}, 575 (2011).

\bibitem{Veldhorst12} 
M. Veldhorst, M. Snelder, M. Hoek, T. Gang, V. K. Guduru, X. L. Wang, U. Zeitler, W. G. v. d. Wiel, A. A. Golubov, and H. Hilgenkamp,
Nature Mat. {\bf 11}, 417 (2012). 

\bibitem{Wang12_STI} 
M.-X. Wang, C. Liu, J.-P. Xu, F. Yang, L. Miao, M.-Y. Yao, C. L. Gao, C. Shen, X. Ma, X. Chen, Z.-A. Xu, Y. Liu, S.-C. Zhang, D. Qian, J.-F. Jia,
and Q.-K. Xue, Science {\bf 336}, 52, (2012). 

\bibitem{Williams12}
J. R. Williams, A. J. Bestwick, P. Gallagher, S. S. Hong, Y. Cui, A. S. Bleich, J. G. Analytis, I. R. Fisher, and D. Goldhaber-Gordon, 
Phys. Rev. Lett. {\bf 109}, 056803 (2012).

\bibitem{Maier12}
L. Maier, J.B. Oostinga, D. Knott, C. Br\"{u}ne, P. Virtanen, G.Tkachov, E.M. Hankiewicz, C. Gould, H. Buhmann, and L. W. Molenkamp, 
Phys. Rev. Lett. {\bf 109}, 186806 (2012).

\bibitem{Cho12} 
S. Cho, B. Dellabetta, A. Yang, J. Schneeloch, Z. Xu, T. Valla, G. Gu, M. J. Gilbert, and N. Mason, Nat. Commun. {\bf 4}, 1689 (2013).  


\bibitem{Oostinga13}
J. B. Oostinga, L. Maier, P. Sch\"uffelgen, D. Knott, C. Ames, C. Br\"une, G. Tkachov, H. Buhmann, and L. W. Molenkamp, 
Phys. Rev. X {\bf 3}, 021007 (2013).


\bibitem{Stanescu10} 
T. D. Stanescu, J. D. Sau, R. M. Lutchyn, and S. Das Sarma,
Phys. Rev. B {\bf 81}, 241310(R) (2010). 

\bibitem{Potter11} 
A. C. Potter and P. A. Lee, Phys. Rev. B {\bf 83}, 184520 (2011). 

\bibitem{Labadidi11} 
M. Lababidi and E. Zhao, Phys. Rev. B {\bf 83}, 184511 (2011). 

\bibitem{Khaymovich11} 
I. M. Khaymovich, N. M. Chtchelkatchev, and V. M. Vinokur,
Phys. Rev. B {\bf 84}, 075142 (2011). 

\bibitem{Virtanen12} 
P. Virtanen and P. Recher, Phys. Rev. B {\bf 85}, 035310 (2012). 

\bibitem{Yokoyama12} 
T. Yokoyama, Phys. Rev. B {\bf 86}, 075410 (2012). 

\bibitem{Black12} 
A. M. Black-Schaffer and A. V. Balatsky, Phys. Rev. B {\bf 86}, 144506 (2012).

\bibitem{GT13b} 
G. Tkachov, Phys. Rev. B {\bf 87}, 245422 (2013).


\bibitem{Fu09} 
L. Fu and C. L. Kane, Phys. Rev. B {\bf 79}, 161408(R) (2009).  

\bibitem{Tanaka09} 
Y. Tanaka, T. Yokoyama, and N. Nagaosa, Phys. Rev. Lett. 103, 107002 (2009). 

\bibitem{Linder10}
J. Linder, Y. Tanaka, T. Yokoyama, A. Sudbo, and N. Nagaosa, Phys. Rev. B {\bf 81}, 184525 (2010); Phys. Rev. Lett. {\bf 104}, 067001 (2010).   

\bibitem{Badiane11}
D. M. Badiane, M. Houzet, and J. S. Meyer,
Phys. Rev. Lett. {\bf 107}, 177002 (2011).

\bibitem{Olund12} 
C. T. Olund and E. Zhao, Phys. Rev. B {\bf 86}, 214515 (2012).

\bibitem{Yamakage12} 
A. Yamakage, M. Sato, K. Yada, S. Kashiwaya, and Y. Tanaka, Phys. Rev. B {\bf 87}, 100510(R) (2013).

\bibitem{Beenakker13b} 
C. W. J. Beenakker, D. I. Pikulin, T. Hyart, H. Schomerus, and J. P. Dahlhaus, Phys. Rev. Lett. {\bf 110}, 017003 (2013).  

\bibitem{Zhang13} 
S.-F. Zhang, W. Zhu, Q.-F. Sun, arXiv:1301.5182

\bibitem{Wieder13}
B. J. Wieder, F. Zhang, and C. L. Kane, arXiv:1302.2113.

\bibitem{Snelder13} 
M. Snelder, M. Veldhorst, A. A. Golubov, and A. Brinkman, Phys. Rev. B {\bf 87}, 104507 (2013).

\bibitem{Potter13} 
A. C. Potter and L. Fu, arXiv:1303.1524.


\bibitem{McMillan68}
W. L. McMillan, Phys. Rev. {\bf 175}, 537 (1968).

\bibitem{Golubov04}
A. A. Golubov, M. Yu. Kupriyanov, and E. Il’ichev, Rev. Mod. Phys. {\bf 76}, 411 (2004).

\bibitem{GT04}
G. Tkachov and V.I. Fal'ko, Phys. Rev. B {\bf 69}, 092503 (2004).

\bibitem{GT05}
G. Tkachov, Physica C {\bf 417}, 127 (2005).

\bibitem{Fagas05}
G. Fagas, G. Tkachov, A. Pfund, and K. Richter, Phys. Rev. B {\bf 71}, 224510 (2005).

\bibitem{Kopnin11}
N. B. Kopnin and A. S. Melnikov, Phys. Rev. B {\bf 84}, 064524 (2011). 

\bibitem{Stanescu11}
T. D. Stanescu, R. M. Lutchyn, and S. Das Sarma, Phys. Rev. B {\bf 84}, 144522 (2011). 


\bibitem{GT11}
G. Tkachov and E. M. Hankiewicz, Phys. Rev. B {\bf 84}, 035444 (2011).

\bibitem{Rohlfing09}
F. Rohlfing, G. Tkachov, F. Otto, K. Richter, D. Weiss, G. Borghs, and C. Strunk,
Phys. Rev. B {\bf 80}, 220507(R) (2009).

\bibitem{Leggett75}
A. J. Leggett, Rev. Mod. Phys. {\bf 47}, 331 (1975). 

\bibitem{Gorkov01}
L. P. Gorkov and E. I. Rashba, Phys. Rev. Lett. {\bf 87}, 037004 (2001).

\bibitem{Santos10} 
L. Santos, T. Neupert, C. Chamon, and C. Mudry,
Phys. Rev. B {\bf 81}, 184502 (2010).

\bibitem{Link_GT13}
This equation was given without derivation in Ref. \onlinecite{GT13}.

\bibitem{Beenakker04}
C.W.J. Beenakker, in {\em Transport Phenomena in Mesoscopic Systems}, edited by H. Fukuyama and T. Ando (Springer, Berlin, 1992).


\end{thebibliography}
\end{document}